lecture1.
\documentstyle[12pt]{report}
\evensidemargin\oddsidemargin
\newcommand{\alt}{\mathrel{\raisebox{-.6ex}{$\stackrel{\textstyle<}{\sim}$}}}
\newcommand{\agt}{\mathrel{\raisebox{-.6ex}{$\stackrel{\textstyle>}{\sim}$}}}

\def\overlay#1#2{\ifmmode \setbox 0=\hbox {$#1$}\setbox 1=\hbox to\wd 0{\hss
$#2$\hss }\else \setbox 0=\hbox {#1}\setbox 1=\hbox to\wd 0{\hss #2\hss }\fi
#1\hskip -\wd 0\box 1}

\def\nv#1 {\noalign{\vskip#1pt}}

\def\abstract#1{\begin{center}\large\bf Abstract\end{center} \vspace{-.135in}
{\narrower\small #1\par}}

\def\gev{{\rm\,GeV}}
\def\tev{{\rm\,TeV}}
\def\L{{\cal L}}
\def\pb{{\rm\,pb}}
\def\fb{{\rm\,fb}}
\def\pbar{\bar p}
\def\tbar{\bar t}

\renewcommand{\labelenumi}{(\alph{enumi})}

\catcode`@=11
\newcount\@tempcntc
\def\@citex[#1]#2{\if@filesw\immediate\write\@auxout{\string\citation{#2}}\fi
  \@tempcnta\z@\@tempcntb\m@ne\def\@citea{}\@cite{\@for\@citeb:=#2\do
    {\@ifundefined
       {b@\@citeb}{\@citeo\@tempcntb\m@ne\@citea\def\@citea{,}{\bf ?}\@warning
       {Citation `\@citeb' on page \thepage \space undefined}}%
    {\setbox\z@\hbox{\global\@tempcntc0\csname b@\@citeb\endcsname\relax}%
     \ifnum\@tempcntc=\z@ \@citeo\@tempcntb\m@ne
       \@citea\def\@citea{,}\hbox{\csname b@\@citeb\endcsname}%
     \else
      \advance\@tempcntb\@ne
      \ifnum\@tempcntb=\@tempcntc
      \else\advance\@tempcntb\m@ne\@citeo
      \@tempcnta\@tempcntc\@tempcntb\@tempcntc\fi\fi}}\@citeo}{#1}}
\def\@citeo{\ifnum\@tempcnta>\@tempcntb\else\@citea\def\@citea{,}%
  \ifnum\@tempcnta=\@tempcntb\the\@tempcnta\else
   {\advance\@tempcnta\@ne\ifnum\@tempcnta=\@tempcntb \else \def\@citea{--}\fi
    \advance\@tempcnta\m@ne\the\@tempcnta\@citea\the\@tempcntb}\fi\fi}
\catcode`@=12

\begin{document}

\thispagestyle{empty}
\setcounter{page}{0}

\font\fortssbx=cmssbx10 scaled \magstep2
\hbox to \hsize{
\hskip.5in \raise.1in\hbox{\fortssbx University of Wisconsin - Madison}
\hfill$\vcenter{\hbox{\bf MAD/PH/780}
            \hbox{July 1993}}$ }

\vspace{1in}

\begin{center}
{\LARGE\bf COLLIDER PHYSICS 1993\footnotemark}\\[.4in]
{\large V.~Barger$^a$  and R.J.N.~Phillips$^b$}\\[.2in]
\it
$^a$Physics Department, University of Wisconsin, Madison, WI 53706, USA\\
$^b$Rutherford Appleton Laboratory, Chilton, Didcot, Oxon OX11 0QX, UK
\end{center}

\footnotetext{Lectures presented by V.~Barger at the {\it VII Jorge Andr\'e
Swieca Summer School}, S\~ao Paulo, Brazil, January 1993.}

\renewcommand{\LARGE}{\Large}
\renewcommand{\Huge}{\Large}

\vfill

\abstract{
These lectures survey the present situation and future prospects in
selected areas of particle physics phenomenology: (1)~the top quark,
(2)~the  Higgs boson in the Standard Model, (3)~strong $WW$ scattering,
(4)~supersymmetry, (5)~the Higgs sector in minimal supersymmetry,
(6)~low-energy constraints on supersymmetry.
}

\vfill
\newpage

\section*{GENERAL INTRODUCTION}
   Our present understanding of elementary particle phenomena is
dominated by the Standard Model (SM), based on three generations of basic
spin-1/2 fermions (quarks and leptons), plus the spin-1 gauge fields of
$\rm SU(3)\times SU(2)_L\times U(1)$ symmetry, plus spontaneous symmetry
breaking by one
doublet of spin-0 scalar fields.  The SM works amazingly well --- all of
its predictions that can be tested have so far been verified to high
precision; no discrepancies have yet been found.  However the predicted
top quark has not yet been confirmed experimentally, nor has the Higgs
boson.  The top quark is not seriously in doubt, since lots of indirect
evidence points to its existence, but it is nevertheless important to
discover it to determine its mass and other properties that enter many
calculations.  The symmetry-breaking sector appears to be much more
arbitrary (one can easily imagine alternatives to the simple Higgs
mechanism) and urgently requires investigation.  This sector can be
probed either by searching for the Higgs boson or by studying the
scattering of longitudinally polarized $W$ or $Z$ bosons; these longitudinal
states arise through symmetry-breaking and their scattering reflects
the underlying mechanism.  These topics are addressed in the first
three lectures: (1)~the top quark, (2)~the standard model Higgs boson,
(3)~strong $WW$ scattering.

   The SM has severe shortcomings however.  It contains
many apparently arbitrary and unrelated parameters; this
arbitrariness might be reduced or explained by some Grand Unified Theory
(GUT), where the different gauge symmetries merge into a single higher
symmetry and the masses may be simply related at a very high energy
scale.  But in plain GUT models associated with electroweak symmetry breaking
an arbitrary fine-tuning of parameters
seems to be needed to prevent the scalar particles from acquiring very large
masses; a search for a solution to this ``hierarchy problem" leads to
Supersymmetry (SUSY), where every fermion has a boson partner and vice versa.
No SUSY partners have yet been discovered but there is encouragement from the
success of SUSY-GUT models, and indeed cosmological Dark Matter
may be due to one of these particles.  SUSY is also a likely ingredient
in an eventual unification of strong and electroweak forces with gravity,
while behind all this there lies perhaps a Superstring theory of everything.
All sorts of new phenomena may arise beyond the SM: extra gauge bosons,
exotic fermions, leptoquarks, nucleon decay, new classes of Yukawa
interactions, etc.  Some of these topics are addressed in the later lectures:
(4)~SUSY and GUTs, (5)~Higgs sector in the minimum SUSY extension of the
SM (MSSM), (6)~low-energy constraints on SUSY.

   We give a selection of references, but make no claim to completeness
and specifically exclude references to standard textbook material.


\chapter*{LECTURE 1:\\ THE TOP QUARK}
\setcounter{chapter}{1}

\section{Introduction}

   The top quark $t$ is an essential part of the third fermion generation
in the SM, together with the $b$-quark, the $\tau$-lepton and its neutrino
$\nu_\tau$.  Their left and right chiral components have the usual
$\rm SU(2)_L\times U(1)_Y$  weak-isospin and hypercharge quantum numbers (with
electric charge $Q=T_3+{1\over2}Y$):
%
\[
\begin{array}{c@{\hspace{3em}}r@{\hspace{3em}}r}
\hline\hline
\nv3
                         &       T_3      &           {1\over2} Y \\
\nv3
\hline
\nv3
\left(\begin{array}{c}
       \nu_\tau    \\
        \tau
\end{array}\right)_{\!\!L} &
\begin{array}{r}
 1\over2    \\
- {1\over2}
\end{array} &
\begin{array}{r}
-{1\over2} \\
-{1\over2}
\end{array}
\\
\nv3
\hline
\nv3
          \tau_R          &
\begin{array}{r} 0 \end{array}      &
\begin{array}{r} - 1 \end{array} \\
\nv3
\hline
\nv3
\left( \begin{array}{c}
         t \\
         b
\end{array} \right)_{\!\!L}   &
\begin{array}{r}
    1\over2   \\
  - {1\over 2}
\end{array}  &
\begin{array}{r}
 1\over 6 \\
 1\over 6
\end{array}
\\
\nv3
\hline
\nv3
           t_R           &
\begin{array}{r}  0 \end{array}       &
\begin{array}{r}  2\over3 \end{array} \\
\nv3
\hline
\nv3
           b_R           &
\begin{array}{r}  0  \end{array}      &
\begin{array}{r} - {1\over3} \end{array} \\
\nv3
\hline\hline
\end{array}
\]

   Here $t$ (and $\nu_\tau$) are the only SM fermions to escape direct
detection.  In the case of $t$, this is apparently because it is very
heavy~\cite{cdftop},
\begin{equation}
          m_t > 108\mbox{ GeV \qquad (1993 SM CDF limit)}\;, \label{m_t CDF}
\end{equation}
compared to other quarks $(m_u = 0.004,\ m_d = 0.007,\ m_s = 0.15,\ m_c = 1.3,\
m_b = 4.8$~GeV).  But we have many indirect indications that top exists:
\begin{enumerate}
\item It is needed to cancel chiral anomalies.

\item It is needed for GIM-suppression of flavor-changing neutral currents.
If $b_L$ had no doublet partner $t_L$, then $BF(B\to \ell^+ \ell^- X) >
0.013$~\cite{barpak}; but the CLEO experimental bound is ${}< 0.0012$ at
90\%~CL~\cite{cleollx}. Also, if $b$ were an SU(2)$_L$ singlet, $B_d^0$-$\bar
B_d^0$  oscillations would be near maximal~\cite{roybb}
but in fact they are not.

\item The $e^+e^- \to \bar bb$ forward-backward asymmetry measures
$T_3 (b_L) - T_3 (b_R)$;  LEP data give a value
$-0.504{+0.018\atop-0.011}$~\cite{lepsm},
confirming the SM value $-1/2$ and showing that $b_L$ belongs to a doublet.

\item With a singlet assignment for $b$, the predicted $Z\to b\bar b$ partial
width is a factor of 15 times smaller than the measured value~\cite{roybb}.
\item Electroweak radiative corrections to all available $Z,\ W$ and deep
inelastic scattering data fit beautifully but require top-quark
contributions with a mass of order~\cite{elfog}
\end{enumerate}
\begin{equation}
m_t = 141 {+17\atop-19} {+17\atop-18} \mbox{ GeV \qquad  (SM electroweak)} \;.
\label{m_t SM}   \end{equation}

\section{Top production, decay and detection}

   Top decays dominantly by  $t \to b W^+$  in the SM (Fig.~1); other
charged-current decays to $dW$ or $sW$ are suppressed by small KM matrix
elements;
neutral-current decays to $uZ$ or $cZ$ are suppressed by the GIM mechanism.
Here $W$ is now known to be on-shell [see Eq.~(\ref{m_t CDF})] and its
subsequent decays are dominantly  $ W^+ \to u\bar d,\ c \bar s,\ \nu \bar e,\
\nu \bar\mu,\ \nu \bar\tau$ with respective branching
fractions 1/3, 1/3, 1/9, 1/9, 1/9, approximately.  The semileptonic modes
$t\to be\nu,\ b\mu\nu$  provide the cleanest signatures, typically containing
\begin{enumerate}
\item an energetic lepton, isolated from jets,
\item missing energy and momentum,
\item a $b$-jet, possibly tagged by a muon from $b\to c \mu \nu$ decay,
\item a displaced vertex from $b$-decay (long lifetime),
\end{enumerate}
which help to separate top events from backgrounds.

\medskip
\begin{center}
\hspace*{0in}

{\small Fig.~1: Top decay in the SM.}
\end{center}

   If $m_t$ is close to threshold, the mean lifetime  $\tau = 1/\Gamma$  is
long enough to allow the usual fragmentation (formation of a hadron)
before the top quark decays. But if  $m_t \agt 120$~GeV the lifetime is
too short to form any hadron (including toponium states) and top
essentially decays as a free quark; in this regime its width is
\begin{equation}
     \Gamma(t \to bW) \sim 0.17 (m_t/M_W)^3 \ \rm GeV  \;, \label{Gamma}
\end{equation}
shown in Fig.~2~\cite{fujii}. $\Gamma_t$ may be measured from the linewidth;
there are also interference effects between gluons radiated from $t$ and $b$,
which depend sensitively on the top lifetime~\cite{khoze}.

\begin{center}
\hspace*{0in}

{\small Fig.~2: SM top decay width versus $m_t$. From Ref.~\cite{fujii}.}
\end{center}

   The only present machine that can now discover SM top quarks is the
Fermilab Tevatron $p\bar p$ collider, with CM energy $\sqrt s=1.8$~TeV.
The lowest-order ($\alpha_s^2$) QCD production subprocesses are light
quark-antiquark and gluon-gluon fusion (Fig.~3).

\begin{center}
\hspace*{0in}

\medskip
{\small Fig.~3: $t\bar t$ production via QCD at the Tevatron.}
\end{center}

\noindent
The total cross section  is
given by a convolution of subprocess cross sections $\hat\sigma$ with parton
distributions of the general form
\begin{equation}
    \sigma(s) = \sum_{i,j} \int dx_1 dx_2 \hat\sigma_{ij}(x_1x_2s,\,\mu^2)
f_i^A(x_1,\,\mu) f_j^B(x_2,\,\mu) \;,  \label{sigma(s)}
\end{equation}
where $A$ and $B$ denote the incident hadrons, $i$ and $j$ are the initial
partons, $x_1$ and $x_2$ are their longitudinal momentum fractions, and
$\mu$ is the renormalization scale. The $t\bar t$ hadroproduction cross section
has been calculated to the next order ($\alpha_s^3$)~\cite{rkellis}; there are
uncertainties from the parton distributions and also from the choice of scale
(Fig.~4). Electroweak subprocesses such as $W^+ g \to t \bar b$, producing
single top quarks, are also interesting but not competitive
at the Tevatron\cite{yuan}.

\begin{center}
\hspace*{0in}

\medskip

\parbox{5in}{\small Fig.~4: Range of predictions for $\sigma(p\bar p\to t\bar
tX)$ at $\sqrt s=1.8$~TeV at order $\alpha_s^3$, with  various scales $\mu$.
{}From Ref.~\cite{berends}.}
\end{center}

   To detect $t\bar t$ production is not easy: it is only a tiny fraction of
the total cross section and fake events (containing leptons and jets)
can arise from relatively copious production of $b$-quarks and $W$ or $Z$.
For $p\bar p$ collisions at $\sqrt s=1.8$~TeV we have~\cite{hmrs}
\[
\vbox{\halign{\tabskip1em
$#\hfil$& \hfil#\tabskip0em& \hfil${}=#\,\rm cm^2$\cr
\sigma(\rm total)& 70 mb& 7\times10^{-26}\cr
\sigma(b\bar b)& 30 $\mu$b& 3\times10^{-29}\cr
\sigma(W)& 20 nb& 2\times10^{-32}\cr
\sigma(Z)& 2 nb& 2\times10^{-33}\cr
\sigma(t\bar t)_{m_t=150}& 10 pb& 1\times10^{-35}\cr}}
\]
The reliability of QCD calculations has been tested by $b$-quark data;
Figure~5 shows that CDF measurements of inclusive $b$-production for
transverse momentum $p_T(b) > p_T^{\rm min}$  approximately agree
with expectations~\cite{hmrs}.

The number of observed events in a given channel is
\begin{equation}
N_{\rm events} = \sigma\times{\rm BF}\times \int\L dt \times\rm efficiency\;.
\end{equation}
Here $\int{\cal L}dt$ is the integrated luminosity, for which the CDF
detector accumulated about 4~pb$^{-1}$ up to 1989. However the D0 detector
is now working too and the accelerator may deliver
25~pb$^{-1}$ per detector in the 1992--3 running, rising to 75~pb$^{-1}$ in
1994 and eventually (with a new main injector) up to 1000~pb$^{-1}$ in 1997. A
possible increase of energy to $\sqrt s=2$~TeV in 1994 would increase
the $t\bar t$ signal by about 30\% for $m_t = 150$~GeV.

\begin{center}
\hspace*{0in}

\parbox{5in}{\small Fig.~5: Inclusive $p\bar p\to bX$ cross sections versus
$p_T^{\rm min}$; preliminary CDF data are compared with QCD expectations. From
Ref.~\cite{hmrs}.}
\end{center}

   Top events have the structure $p\bar p\to t\bar t\to (bW^+)(\bar bW^-)$
where typically the $b$-quarks appear as jets and each $W$-boson appears either
as an isolated lepton (plus invisible neutrino) or as a pair of quark
jets. However, the various partons can also radiate additional gluons or
quarks, and final state partons can overlap, so the net number of jets is
not fixed.  Typical topologies for $t\bar t$ signal and $b\bar b$ background
events are shown in Fig.~6.  The $b\bar b$ background process produces a
lepton in or near a jet (i.e.\ non-isolated) and can be greatly suppressed
by a stringent isolation requirement.  The other major background, from
$W+{}$QCD~jets, can give isolated leptons but usually gives less central jet
activity than the $t\bar t$ signal; furthermore the jets usually do not
contain a $b$-jet.

\begin{center}
\hspace*{0in}

{\small Fig.~6: Topologies of typical $t\bar t$ signal and $b\bar b$ background
events.}
\end{center}

\section{Top search strategies}

   Exploiting the leptonic $W$-decays, one usually requires either one
or two isolated leptons, for which $t\bar t$ states have the following
branching fractions:
\begin{eqnarray}
     B(t\bar t \to e \mu X)        &=&  0.024 \;, \nonumber\\
     B(t\bar t \to ee \mbox{ or } \mu\mu X)  &=&  0.024 \;,  \label{BFs}\\
     B(t\bar t \to e \mbox{ or } \mu +jets) &=&  0.29  \;. \nonumber
\end{eqnarray}
There are then various strategies, based on detecting
\begin{enumerate}
\item Dileptons: this is the smallest but cleanest channel. We require two
isolated opposite-sign leptons, not back-to-back, plus missing-$p_T$ (denoted
$\overlay/p_T$). The $b\bar b$ background is suppressed (mostly non-isolated
and back-to-back in azimuth);
$W+{}$jets does not contribute; Drell-Yan and $Z+{}$jets is suppressed by
$\overlay/p_T$
(and additionally if we restrict to $e\mu$ cases). A small background from
direct $WW$ and $WZ$ production remains, further suppressed by requiring
extra jets.

\item Single lepton plus jets: this is bigger but dirtier. We require one
isolated lepton plus large $\overlay/p_T$ plus several jets, of which two have
invariant mass $m(jj)\sim M_W$.  Then $b\bar b$ is suppressed but $W+{}$jets is
only partly suppressed.  This signal is bigger [see Eq.~(\ref{BFs})] and also
offers a complete
reconstruction of top from 3 of the jets  (unlike (a) where there is always
a missing neutrino), but the $W+{}$jets background remains problematical.

\item $b$-tags.  The presence of a muon near a jet, with $p_T > 1$~GeV
relative to the jet axis, tags it as probably a $b$-jet.  Also the
presence of a displaced vertex (with suitable conditions on the
emerging tracks) can give an efficient $b$-tag.  Adding such a tag would
purify both signals (a) and (b) above, but at some cost to the signal
event rate --- perhaps a factor 10 for muon-tagging or a factor 3 for
vertex-tagging.   CDF have used muon-based $b$-tagging to sharpen up
their top search in the single-isolated-lepton channel; a vertex-tagger
is now in use too. A recent study of $b$-tagging in the heavy-top search is
given in Ref.~\cite{han}
\end{enumerate}

   We can illustrate this discussion with some numbers, assuming
integrated luminosity 100~pb$^{-1}$, acceptance cuts $p_T > 15$~GeV on each
lepton and jet and $\overlay/p_T$, plus further detection efficiency
factors 50\% for each lepton. Predicted numbers of events are
then~\cite{protopop}
\[
\begin{array}{c@{\qquad}c@{\qquad}c@{\qquad}c} \hline\hline
m_t\rm\ (GeV) & \sigma\rm\ (pb) & e + \mu (+2\rm\,jets) &
 e\hbox{ or }\mu > 2\,\rm jets \\  \hline
100 &  88 & 50\ (30) & 570 \\
120 & 34 & 20\ (16) & 300 \cr
140 & 15 & 10\ (9) & 160\cr
180 & 3 & 2\ (2) & 40\cr  \hline
\rm backgrounds \cr
Z\to \tau\tau & 200 & 30\ (7) & \cr
W\to\ell\nu & 4400 & & 150 \cr \hline\hline
\end{array}
\]

Comparing this with the current Tevatron situation (luminosity
${}\approx 20$~pb$^{-1}$) we would expect about 4(2) $e\mu$ candidate events
for $m_t=120(140)$~GeV, in the range preferred by SM theory Eq.~(\ref{m_t SM}).
In fact there are three candidate $t\bar t\to{}$dilepton events from CDF and
one from D0, so the numbers are not inconsistent with expectations for $m_t\alt
160$~GeV.

   Figures 7--10 illustrate important aspects of the dilepton search
strategy~\cite{bbp2}. Figure~7 shows how a missing-$p_T$ cut (in this case
${\not\hskip-1.5pt p}_T > 20$~GeV) discriminates strongly against $\gamma^*\to
\ell^+\ell^-$  and $Z\to\ell^+\ell^-$  backgrounds.  Figure~8 shows how an
azimuthal angle-difference cut  $30^\circ <  \Delta\phi(\ell^+\ell^-)  <
150^\circ$  discriminates against $b\bar b\to\ell^+\ell^-$  backgrounds.
Figure~9 shows how the number of accompanying jets in
$t\bar t\to \ell^+\ell^-$  events changes with $m_t$; we see that in the
neighborhood  $m_t\sim M_W + m_b$  the recoiling $b$-quarks are too soft to
form jets so the multiplicity falls, and it is not efficient to
demand jets here. Fortunately the $t\bar t$ signal here is much stronger
than the remaining backgrounds and we do not need help.  At higher $m_t$
values where the signal falls and the $WW$ background becomes a problem,
the $b$-jets are hard and can be used for additional discrimination.
Finally, Fig.~10 shows that the $t\bar t\to \ell^+\ell^-$  signal can be
distinguished for higher masses $m_t = 150$--200~GeV and even beyond, by
making a more severe requirement $p_T > 30$~GeV  on the two
accompanying jets; the $b$-jets from $t\to bW$ are naturally very hard in this
mass range so there is little cost to the signal, but the $WWjj$ background
goes down by a factor~4.  The good news in all this is that the
``gold-plated" dilepton signal remains essentially background-free
through the whole of the SM-favored mass range and well beyond.

\begin{center}
\hspace*{0in}

\medskip

\parbox{5.5in}{\small Fig.~7: $\gamma^*$ and $Z$ backgrounds to the top
dilepton signal are suppressed by a cut $\overlay/p_T>20$~GeV~\cite{bbp2}.}

\bigskip

\hspace*{0in}

\parbox{5.5in}{\small Fig.~8: The $b\bar b$ background to the top dilepton
signal is suppressed by rejecting back-to-back leptons (in azimuth), with a cut
$30^\circ<\Delta\phi<150^\circ$~\cite{bbp2}.}

\bigskip\bigskip

\hspace*{0in}

\parbox{5.5in}{\small Fig.~9: Topological cross sections for dileptons plus $n$
jets, versus $m_t$, at the Tevatron (detection efficiency factor $\sim0.3$ not
included)~\cite{bbp2}.}

\bigskip

\hspace*{0in}

\parbox{5.5in}{\small Fig.~10: $p\pbar \to\mbox{dilepton + 2-jet event rates at
}\sqrt s = 2$~TeV, for $t\tbar$ signal and $WWjj$ background. Jet cuts
$p_T(j) > 15$, 30 GeV are compared~\cite{bbp2}.}

\end{center}

   We turn now to single-lepton search strategies, seeking  $t\bar t$
events with one $t\to bW\to b\ell\nu$  plus one  $t\to bW \to bjj$  decay.
We start simply by looking for  $W\to \ell\nu$  events, requiring a hard
isolated lepton plus substantial  $\overlay/p_T$;  the distinctive signature
of $W\to \ell\nu$ is that the ``transverse mass"  $m_T$  peaks sharply near
$M_W$. Here $m_T$ is defined by
\begin{equation}
  m_T^2(e\nu) = (p_{Te}+\overlay/p_T)^2 - (\vec p_{Te}+\vec{\overlay/p}_T)^2 =
2p_{Te}\overlay/p_T(1-\cos\phi_{e\nu})   \label{m_T^2}
\end{equation}
and its peak is a kinematical property, related to the ``Jacobian" peaks
of  $p_T(e)$  and  $p_T(\nu)\simeq\overlay/p_T$, but with the special virtue
that it
is much less smeared by transverse motion of the $W$.  For on-shell
$W$-decays the shape of the  $m_T$  distribution is predictable.  If $m_t
< M_W + m_b$,  $t\bar t$ events would contribute off-shell $W$-decays that
would distort the shape in a characteristic way; Figs.~11 and 12 show
theoretical examples and some real data.
Since the  $t\bar t$ signal has
typically several accompanying jets while the background from plain
$W$-production has typically little jet activity, the signal becomes
more striking when we require more jets (Fig.~11); it should be
detectable in events with  $n \ge 2$ jets~\cite{bbp2}.  Figure~12 shows
comparisons of theoretical distributions with CDF 2-jet and 1-jet data; the
absence of any detectable signal in the early CDF data gave a limit  $m_t >
77$~GeV at 95\%~CL.

   If however $m_t > M_W + m_b$,  the $W$ is mostly on-shell and no
distortion of the  $m_T$  distribution is expected.  In this case we must
simply look for an excess in  $W + n$-jets  events, compared to the
background from plain $W$ production.  This background is huge.  To
reduce it we first require that $n$ be large, say $n \ge 3$  or
$n \ge 4$.  We can also require that two of the jets have invariant
mass $m(jj) \simeq M_W$  (since the signal contains $W \to jj$), but
with experimental uncertainties and many possible jet pairings this
condition is not very stringent.   It is much more effective to
require that one jet is $b$-tagged, as shown in Fig.~13; the cross sections
shown here do NOT include the tagging efficiency, which depends on
experimental details but is of{\parfillskip0pt\par}

\begin{center}
\hspace{0in}

\medskip

\parbox{5.5in}{\small
Fig.~11: Theoretical examples of single-lepton signals in the $m_T$
        distribution, for $m_t < M_W+m_b$; $n$ denotes jet
        multiplicity~\cite{bbp2}.}

\bigskip

\hspace{0in}

\medskip

\parbox{5.5in}{\small
Fig.~12: Single-lepton transverse mass distributions at the Tevatron with
      (a)~2 jets and (b)~1 jet. Solid curves are the calculated $W+{}$jets
      background, the dashed curve is the expected signal for
      $m_t=70$~GeV~\cite{cdf1}.}

\end{center}

\noindent
order 30\% for CDF vertex-tagging. We see that,
after $b$-tagging and requiring $n \ge 3$, the $t\bar t$ signal greatly
exceeds the $W+{}$jets backgrounds through the range $m_t = 95$--170~GeV
or so.  The signal diminishes below 95~GeV because the $b$-jets get
soft.

\begin{center}
\hspace{0in}

\parbox{5.5in}{\small
Fig.~13: Effects of $b$-tagging on the single-lepton $t\bar t$ signal at the
      Tevatron, for various jet multiplicities~\cite{protopop}. Jets and
      leptons are required to have $p_T(j)>15$~GeV, $p_T(l)>20$~GeV,
      $|\eta(j)| < 2,\ |\eta(l)| < 1$.}
\end{center}

   With an eventual luminosity 1000~pb$^{-1}$, the expected event yields at
the Tevatron depend on $m_t$ as follows~\cite{protopop}:
\[
\begin{array}{c@{\qquad}c@{\qquad}c}
m_t\ \rm (GeV) & e\hbox{ or }\mu+4\,\rm jets\ events & e+\mu\ \rm events\\
120 & 1380 & 240 \\
140 & 850 & 98 \\
180 & 260 & 24 \\
210 & 140 & 12 \\
240 & 60 & 5
\end{array}
\]
The mass ranges where a firm signal could be established, and the lower
bound that could be set in the case of no signal, depend on luminosity
like this~\cite{protopop}:
\[
\begin{array}{c@{\qquad}c@{\qquad}c}
\L & \hbox{claim discovery} & \hbox{90\% CL bound}\\
100\pb^{-1} & m_t\leq 150\gev & {}>180\gev \\
1000\pb^{-1} & m_t\leq 220\gev & {}>250\gev
\end{array}
\]

\section{Further considerations}

   When a top signal is found, $m_t$ can be estimated from various
dynamical distributions that are sensitive to it~\cite{bbp2}, {\it e.g.}
\begin{enumerate}
\item invariant masses $m(\ell_1,\ell_2)$ of two isolated leptons;
\item invariant mass $m(\ell,\mu)$ of single isolated lepton and
(opposite-sign) muon tagging the associated $b$-jet;
\item invariant mass $m(jjj)$ of three jets accompanying single lepton;
\item cluster transverse masses such as $m_T(\ell\ell jj, \overlay/p_T)$;
\item variables arising in families of explicit event reconstructions.
\end{enumerate}
Maximum-likelihood methods can be used too~\cite{kondo,dalitz}. With
1000~pb$^{-1}$ of luminosity, the Tevatron could determine $m_t$ to 5~GeV or
better (at least up to about 170~GeV).

   For the planned $pp$ supercolliders SSC ($\sqrt s=40$~TeV) near Dallas
and LHC ($\sqrt s=15.4$~TeV originally, now 14~TeV) at CERN,  $t\bar t$
production will be
enormously larger than the Tevatron rates.  Both the intrinsic cross
sections (Fig.~14) and the planned luminosities are much bigger.  If
$m_t=150$~GeV the cross sections and event rates will be
\begin{eqnarray}
\mbox{SSC: } \sigma(pp\to t\bar tX)&=&12\mbox{\,nb\quad giving }1.2\times10^7
\,\mbox{events/year} \;,\nonumber \\
\mbox{LHC: } \sigma(pp\to t\bar tX) &=&\phantom02\mbox{\,nb\quad giving }
2.0\times10^7
\,\mbox{events/year} \;. \nonumber
\end{eqnarray}
One expects to measure $m_t$ to 2--3~GeV using the distribution $m(\ell,\mu)$
of isolated lepton plus tagging muon described in (b) above; since both
$\ell$ and $\mu$ originate from the same parent $t$, their invariant mass
distribution (Fig.~15) depends only on the decay mechanism.

\begin{center}
\hspace{0in}

{\small Fig.~14: $t\bar t$ production at the Tevatron, LHC and
SSC~\cite{reya}.}

\hspace{0in}

{\small Fig.15: $m_t$ dependence of $m(\ell,\mu)$ distribution~\cite{massem}.}
\end{center}

   So far we have assumed purely $t\to bW$  SM decays, but other modes
are possible beyond the SM.  If there exist charged Higgs bosons $H^\pm$,
then decays like
\begin{equation}
          t \to b H^+\to b c \bar s,\, b \nu \bar\tau \;,
\end{equation}
become possible.  If the competing $t\to bH^+$  mode is strong it will reduce
the SM $t\to bW$  branching fraction, reducing the SM signals we have
discussed.  In return we get some new signals: the $b c \bar s$
 final state is similar
to one of the SM modes but with a different mass peak at $m(c\bar s)=m_{H^+}$;
the $b \nu \bar\tau$ mode can be recognized by an excess of $\tau$ over $e$ or
$\mu$ production (lepton non-universality).  In models with two Higgs doublets
the various branching fractions are controlled by a parameter
$\tan\beta$, the ratio of the two vevs, constrained to lie in a range $0.2 \alt
\tan\beta \alt
100$ if the couplings are perturbative~\cite{bhp}; Fig.~16 shows how the SM and
Higgs modes compete and combine to
give a range of possible  $t\to be\nu$ and $t\to b\tau\nu$  fractions for an
example with $m_t=150,\ m_{H^+}=100$~GeV.  At large $\tan\beta$ a dramatic
excess
of $\tau$ over $e$ or $\mu$ is predicted; but at small $\tan\beta$ both the SM
signatures based on $t\to e(\mu)$ and the new signature based on $t\to \tau$
are strongly suppressed and top would be very difficult to discover at hadron
colliders~\cite{bp1}.

   In SUSY models there may be other new modes such as
\begin{equation}
    t \to \tilde t_1 \tilde Z_i\to c \tilde Z_1 \tilde Z_i \; ,
\end{equation}
where $\tilde t_1$ is the light squark partner of $t$,  $\tilde Z_i$ are
neutralinos (with
$\tilde Z_1$ the lightest), if the SUSY particles are light enough.  These
too would compete and deplete the SM signals; a top quark as light as
65~GeV, with dominant SUSY decays, is not yet ruled out by Tevatron
data~\cite{bt}.

\begin{center}
\hspace{0in}

\parbox{5.5in}%
{\small Fig.~16: Sample $\tan\beta$ dependence of $t$ and $H^+$ branching
fractions\cite{bp1}. The hidden top region where all semileptonic signals are
suppressed is $\tan\beta\alt0.3$.}
\end{center}

   Finally we come to the possibilities at future $e^+e^-$ linear
colliders, with luminosities of order 20~fb$^{-1}$/year.  Figure~17 shows
cross sections versus CM energy relative to the $e^+e^-\to\mu^+\mu^-$ cross
section; the kink in the $e^+e^-\to \bar qq$ rate is due to $\bar tt$
production (here assuming $m_t=150$~GeV).  At $\sqrt s=500$~GeV the $e^+e^-\to
t\bar t$ event rate would be around $10^4$/year, comparable with the Tevatron
rather than the SSC; however the events would be much cleaner and top
parameters would be easier to extract.  An $m_t$ measurement with
statistical uncertainty 0.3~GeV from 10~fb$^{-1}$ luminosity is
expected~\cite{kuhn}.  The width $\Gamma_t$ could also be accurately measured
near the threshold energy, either from the energy-dependence of the
cross section (Fig.~18), or from the momentum spectrum of $t$, or from a
forward/backward asymmetry~\cite{fujii,kuhn,peskin}.

\begin{center}
\hspace{0in}

{\small Fig.~17: Cross sections for possible high-energy $e^+e^-$
        colliders~\cite{burke}.}

\bigskip\bigskip

\hspace{0in}

{\small Fig.~18: $\sigma(e^+e^-\to\bar tt)$ near threshold~\cite{kuhn}.}
\end{center}


\renewcommand{\chapter}{\section}


\end{document}